\documentclass[aps]{revtex4}
\usepackage{graphicx} 
\usepackage[latin1]{inputenc}

\usepackage{amsmath}
\usepackage{amsfonts}
\usepackage{amssymb}

\def\gsim{\:\raisebox{-0.5ex}{$\stackrel{\textstyle>}{\sim}$}\:}
\begin{document}
\title {The Need for New Search Strategies for Fourth Generation Quarks at the LHC}
\author{Michael Geller$^1$}

\author{Shaouly Bar-Shalom$^1$}
\author{Gad Eilam $^{1,2}$}
\affiliation{
\begin{enumerate}
\item Physics Department, Technion-Institute of Technology, Haifa 32000, Israel.
\item Physics Department,University of Virginia, Charlottesville VA 22904,
USA.
\end{enumerate}
}

\begin{abstract}

Most limits on the fourth generation heavy top quark (the $t^\prime  $) are based
on the
assumed dominance of $t^\prime   \to Wb$, which is expected to be case in the
minimal fourth generation framework with a
single Higgs (the so called SM4). Here we show, within a variant of a Two
Higgs Doublet
Model with four generations of fermions, that, in general, a different
$t^\prime  $ detection
strategy is required if the physics that underlies the new heavy fermionic
degrees of freedom goes
beyond the "naive" SM4. We find that the recent CMS lower bounds: $m_{t^\prime}< 450$ GeV in the semi-leptonic channel $pp\rightarrow t^\prime  \overline{t^\prime  } \rightarrow \ell \nu bq\overline{q} \overline{b} $ and $m_{t^\prime}< 557$ GeV in the dilepton channel $pp\rightarrow t^\prime\overline{t^\prime}\rightarrow \ell^+\ell^- \nu \overline{\nu} b\overline{b}$, that were obtained using the customary (SM4-driven) detection strategies, do not apply.

In particular, we demonstrate that if the decay $t^\prime \to ht $ dominates, then applying the "standard" CMS search tools leads to a considerably relaxed lower bound: $m_{t^\prime} \gsim 350$ GeV. We, therefore, suggest an alternative search strategy that is more sensitive to beyond SM4 dynamics of the fourth generation fermions.
\end{abstract}

\maketitle
\section{Introduction}
The origin of the ElectroWeak Symmetry Breaking (EWSB), one of the most 
studied theoretical puzzles in particle physics, is yet unknown. 
It is anticipated that the LHC, that has already began its historic 
mission of collecting data at 7 TeV collision energy, will provide 
insight into the heart of this problem. In the Standard Model (SM), EWSB is 
triggered by the Higgs mechanism with a single fundamental scalar, 
leaving its mass unprotected from radiative corrections at the cutoff scale. 
This is known as the hierarchy problem: the tension between the cutoff scale 
(Planck scale or GUT scale) and the seemingly unnatural ElectroWeak (EW) scale, which 
is usually interpreted as evidence for new physics at the TeV scale. 

Hints for the existence of a Higgs boson have been recently reported at
the LHC: ATLAS finds an excess of events around $M_h=126$ GeV with a local significance of 3.5 standard deviations \cite{ATLASHiggs}, while CMS sees a smaller excess with a local significance of 3.1 sigma around $M_h=124$ GeV \cite{CMSHiggs}.
Later, it was shown at the Tevatron \cite{TevatronHiggs} that there is 
a possibility of an excess of events corresponding to $M_h$ between
115 GeV and 135 GeV. It is, therefore, tempting to consider a Higgs mass
in the range $(120-130)$ GeV.  
The LHC with higher integrated 
luminosity and 8 TeV c.m. energy should unveil
more precise details about the Higgs boson. 

In the past few years, several tensions between experiments and the SM predictions were observed in flavor physics \cite{sl07,sl08,lenz1,bona,slprl10,sl10,hfag10,d0dimuonprl,CMSLHCb-mumu}. It was noted, that adding a fourth sequential generation  (for reviews see \cite{sher0,hou2009-rev,SM4proc,khlopov}) to the SM (the so called SM4 which in some instances will be referred to as the "naive" four generations model)
might not only be able to resolve these tensions \cite{SAGMN08,SAGMN10,ajb10B,buras_charm,gh10,NS3,lenz_fourth11,lenz_fourth12,Alok:2010zj,BNS,Buras_Recent,Wingerter,godbole1}, but may also address the hierarchy problem \cite{DEWSB,holdom-new,hung-new}. In particular, when the new fermion states are heavy enough, their Yukawa couplings are driven to a Landau pole or a fixed point, possibly at the TeV scale, where some form of strong dynamics might be responsible for dynamically breaking the EW symmetry. Besides, the additional CP-violating phases in the presence of a fourth family \cite{jarlskog} may be useful for Baryogenesis \cite{fok,gh2011,gh08}.

The fact that adding a heavy fourth generation to the SM implies new physics at ${\cal O}\left({\rm TeV}\right)$, serves as a strong hint that the SM4 is not an appropriate framework to account for the expected (possibly strong) dynamics in the presence of these new heavy fermionic states. For this purpose and others, models that go beyond the SM4 (BSM4) were introduced as an alternative to the naive SM4 framework \cite{Luty,sher1,hung,hashimoto1,hashimoto2,wise1,FermCond,SBS,Warp4,Warp4_Burd,AFB,MSSM4,yue}.

 One way to incorporate the low-energy effects of such possible TeV-scale strong dynamics in an acceptable phenomenological approach is multi-Higgs scenarios, which may naturally account for the composite scalars associated with the new strong dynamics, see e.g., \cite{Luty,sher1,hung,hashimoto1,hashimoto2,wise1,FermCond,SBS}.  An example of a model designed for this purpose, the 4G2HDM (a model with 2 Higgs doublets and a fourth family of fermions), was proposed in \cite{SBS}, and was shown to agree with Electroweak Precision Data (EWPD) in a wide range of its parameter space. An alternative interesting approach to the compositeness picture is warped extra-dimension scenarios, see e.g., \cite{Warp4_Burd,Warp4}. Other examples of BSM4 scenarios were suggested in \cite{MSSM4,AFB,Chen:2012wz,godbole2}. 

Recently, assuming the SM4 framework,  CMS reported \cite{CMS_limit,CMS_limit_dilepton} a 450 GeV lower limit on the $t^\prime  $ mass in the semileptonic channel ($pp\rightarrow t^\prime  \overline{t^\prime  }\rightarrow \left[W^+\right]_{hadronic} b \left[W^-\right]_{leptonic} \overline{b}\rightarrow \ell \nu bq\overline{q} \overline{b} $), and a 557 GeV lower limit in the dilepton channel ($pp\rightarrow t^\prime\overline{t^\prime}\rightarrow \left[W^+\right]_{leptonic} b \left[W^-\right]_{leptonic} \overline{b}\rightarrow \ell^+ \ell^- \nu \overline{\nu} b\overline{b}$), replacing the weaker Tevatron limits \cite{CDF_limit,D0_limit}. The most recent lower bounds on the $b^\prime$ mass are 480 GeV \cite{Atlas_bprime} (Atlas) and 611 GeV \cite{CMS_bprime} (CMS). In addition, the single Higgs of the SM4 is excluded now by CMS in the range $120-600$ GeV \cite{CMS_H_limit}. We argue, that these seemingly stringent limits apply only to the SM4 framework, which is disfavored for reasons that are outlined above. In particular, in BSM4 frameworks, collider signals of fourth generation fermions and/or Higgs particles may be drastically altered, leading to weaker bounds. An example of this was discussed in \cite{Val4}, where an extended Higgs sector (with 4th generation fermions) may highly suppress the signal for Higgs production, circumventing the current limits on the SM4 Higgs mass. 

The purpose of this work is, therefore, to illustrate that, in general, the limits on the $t^\prime  $ mass can be considerably weaker with respect to the SM4 case. We demonstrate that in the 4G2HDM framework of \cite{SBS}, where new decay patterns of $t^\prime  $ emerge through the scalar sector: e.g., $t^\prime  \rightarrow ht$, $t^\prime   \rightarrow H^+b$. We thus propose an alternative search method, more general and, therefore, more suitable to searches for BSM4 scenarios.

\section{$t^\prime  $ detection at the LHC- beyond SM4}

 As mentioned above, CMS has recently reported \cite{CMS_limit} a 
557 GeV lower bound on the $t^\prime  $ mass, replacing the previous limits  by 
CDF \cite{CDF_limit} (358 GeV) 
and D0 \cite{D0_limit} (285 GeV) \cite {ATLAS_limit}. 
A summary on the Tevatron limits can be found at \cite{Ivanov}. These searches assume $Br\left(t^\prime   \rightarrow W^+b\right) \sim {\cal O} \left(1 \right)$ (SM4) and focus on either the semileptonic channel ($1\ell+nj+\rlap{ /}{E}_T$) or the dilepton channel ($2\ell+nj+\rlap{ /}{E}_T$) in the $t^\prime  $ pair production (for a limit that takes into account the branching ratios in $t^\prime  $ decay within the SM4, see \cite{SBS_Wh}):
\begin{equation}
pp\rightarrow t^\prime  \overline{t^\prime  }\rightarrow \left[W^+\right]_{hadronic/leptonic} b \left[W^-\right]_{leptonic} \overline{b}\rightarrow l \nu bq\overline{q} \overline{b}/l^+l^- \nu \overline{\nu} b \overline{b} 
\end{equation}

The limits for the semileptonic channel were extracted using two dimensional fits to $M_{fit}$ and $H_T$ - the reconstructed mass of the $t^\prime  $ and the scalar sum of $p_T$ of the visible jets and leptons and missing $E_T$, respectively. In particular, the mass of the $t^\prime  $ is reconstructed by fitting each event to the process $pp\rightarrow W^+W^-b\overline{b}\rightarrow \ell \nu bq\overline{q} \overline{b}$, i.e., taking the 4 leading jets (with highest $p_T$) and minimizing the $\chi ^2$ for the equations: 
\begin{equation}
\begin{split}
m\left(l\nu\right)&=M_W\\
m\left(q\overline{q}\right)&=M_W\\
m\left(l\nu b\right)&=m\left(q\overline{q}b\right)\equiv M_{fit}
\end{split} \label{CMS_Method}
\end{equation}

On the other hand, the analysis for the dilepton channel relies on the fact that $M_{lb}$, which is the invariant mass of a pair of any lepton and a $b$-jet in the event, is much higher in the $t^\prime  \overline{t^\prime  }$ signal with respect to the leading $t\overline{t}$ background. In particular, in the case of $t\overline{t}$, $M_{lb}$ has an upper bound that corresponds to the mass of the top quark, and therefore in the region above 170 GeV ("signal region") $M_{lb}$ is a clean signal for the $t^\prime  $.

\begin{figure}
\begin{center}
	\includegraphics[scale=0.3]
{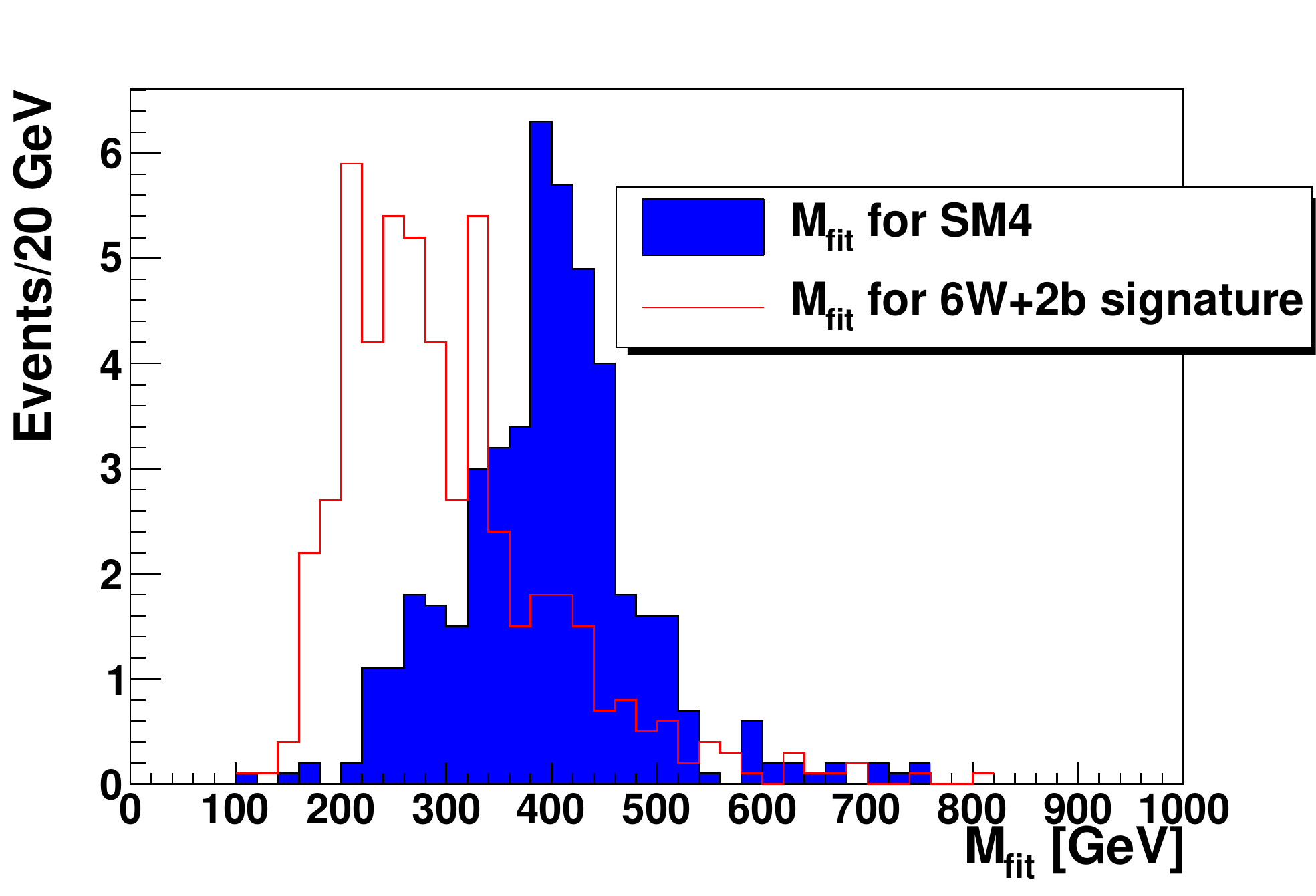}
\end{center}
\caption{$M_{fit}$ distribution for the SM4 $2W+2b \to l\nu \overline{b} b \overline{q} q$ signature (blue) and for the 4G2HDM 6$W$+2$b$ signature (red),  for a set of 7 TeV LHC events with $\int Ldt=1$ fb$^{-1}$. For both signatures  
$m_{t^\prime  }=450$ GeV is assumed. The peak of the distribution of $M_{fit}$ for the SM4 signature is around $m_{t^\prime  }$, while for the new signature the peak is shifted to a significantly lower value coinciding with the peak of the $\overline{t}t$ background.
\label{MFit_ww}}
\end{figure}

 Below we will study the relevance of these limits in the general BSM4 case, in the feasible scenario where $t^\prime  \rightarrow W^+b $ is not the leading $t^\prime  $ decay channel. This may occur in models with an extended Higgs sector, e.g.  in the 4G2HDM suggested in \cite{SBS}. In such models, the new scalars play a crucial role in the $t^\prime  $ phenomenology, and new channels like $t^\prime  \rightarrow ht$, $t^\prime  \rightarrow H^+b$ (see also \cite{sher1}) or even the SM4 forbidden $t^\prime  \rightarrow W^+b^\prime  $  may dominate. In \cite{SBS} it was shown that all these channels are allowed by EWPD, and some of them dominate within large areas in the (EWPD allowed) parameter space.  
 
  Thus, in this scenario, $pp\rightarrow \ell \nu bq\overline{q} \overline{b}/\ell^+\ell^- \nu \overline{\nu} b \overline{b}$ are no longer the leading signatures for $t^\prime  $ pair production at the LHC; the new channels can dominate, giving rise to several possible new signatures for $t^\prime  $ at the LHC (see section 4). We argue that for these cases, the CMS fit to $pp\rightarrow \ell \nu bq\overline{q} \overline{b}$ and $pp\rightarrow \ell^+\ell^- \nu \overline{\nu} b \overline{b}$ should fail, as it attempts to impose these specific SM4-motivated dynamics on processes that have a completely different topology. In particular, for the semileptonic channel, taking only 4 leading jets out of all the jets in the event will miss a large part of the energy of the event and will, therefore, result in $M_{fit}$ being substantially lower- exactly where the main $t\overline{t}$ background is. An example of that can be seen in Figure~\ref{MFit_ww} for the 6$W$+2$b$ signature which occurs in the 4G2HDM (see section 4). For the dilepton channel, the analysis will fail for signatures with more than 2 leptons or $b$-jets, as the combinatorial background will lower the $M_{lb}$ (see section 6), resulting in much less events in the signal region.
  
 In the semileptonic channel the kinematics of each event is completely determined, thus we suggest using a reconstruction that does not assume a specific decay chain, and instead reconstructs any event of the type $1\ell+nj+\rlap{ /}{E}_T$ under the sole assumption that the lepton and the missing energy are the result of the leptonic decay of a W boson, similarly to \cite{SGA}. This will result in the correct reconstruction of $t^\prime  \overline{t^\prime  }$ events (see section 5), regardless of the specific decay signature. 
 
 The specifics of the such a reconstruction strategy are the following: we select all the jets with transverse momentum that is high enough, we then reconstruct the leptonic W and choose the correct partition of the event (which jet is originated from $t^\prime  $ and which one from $\overline{t^\prime  }$ ) by minimizing the $\chi ^2$ of the following equations
 \begin{equation}
 \begin{split}
 m\left(l\nu\right)&=M_W \\
m\left(Left\;  Side \right)&=m\left(Right\; Side \right)\equiv M_{gen} 
\end{split} \label{Our_Method}
 \end{equation}

Another important point to notice is that for all the new signatures, the number of jets in the final state is higher than the SM background, and for some of the signatures there is also an excess of $b$-jets, as will be shown in section 4. We will use cuts based on these properties in section 5, and, as we will further show, this will result in higher detection sensitivity for these signatures.

\section{The 4G2HDM}
Let us recapitulate the underlying properties of the 4G2HDM, the full account of which appears in ~\cite{SBS}.
\\ The 4G2HDM is a two-Higgs doublet model with 4 generations of chiral fermions, where the Yukawa terms are defined as follows:
\begin{equation}
\begin{split}
L_{Y}&=-\overline{Q}_{L}\left(\Phi_{l}F\left(I-I_{d}^{\alpha_{d}\beta_{d}}\right)+ \Phi_{h}F\cdot I_{d}^{\alpha_{d}\beta_{d}}\right)d_{R}-
\overline{Q}_{L}\left(\tilde{\Phi}_{l}G\left(I-I_{u}^{\alpha_{u}\beta_{u}}\right)+\tilde{\Phi}_{h}G\cdot I_{u}^{\alpha_{u}\beta_{u}}\right)u_{R}+h.c.
\end{split}
\end{equation}
such that $Q_L$ is the left-handed SU(2) quark doublet, $f_{L(R)}$ are the left (right)- handed fermion fields and $F$,$G$ are arbitrary 4x4 Yukawa matrices in flavor space. Also, $I$ is the identity matrix, $I_{q}^{\alpha_{q}\beta_{q}}$ are diagonal matrices defined as $I_{q}^{\alpha_{q}\beta_{q}}=diag\left(0,0,\alpha_q,\beta_q\right)$ and 
$\Phi_{l}$, $\Phi_h$ are the two Higgs doublets.
\\The different choices of $\alpha_q,\beta_q$ will result in different Yukawa textures and, therfore, different models.
In this work we assume $\left(\alpha_d,\beta_d,\alpha_u,\beta_u\right)=\left(0,1,0,1\right)$ for which $\Phi_h$ couples only to the $4^{th}$ generations quarks, while $\Phi_l$ couples to the first 3 generations. This setup was defined as 4G2HDM of type I in \cite{SBS} and is partly motivated by the absence of (tree-level) FCNC in the first 3 generations. In particular, in this case, FCNC is only allowed between the third and fourth generations and it plays a crucial role in the phenomenology of the 4th generation fermions.
\\The physical Higgs scalars that emerge after EWSB are: $H^{\pm}$ (charged scalar),$h$,$H$ (lighter and heavier CP-even neutral scalars) and $A$ (CP-odd neutral scalar). Their couplings to the quarks are given by:
\begin{equation}
\begin{split}
L(hq_{i}q_{j})&=\frac{g}{2M_{W}}\bar{q}_{i}\left(m_{q_{i}}\frac{s_{\alpha}}{c_{\beta}}\delta_{ij}-\left(\frac{c_{\alpha}}{s_{\beta}}+\frac{s_{\alpha}}{c_{\beta}}\right)\left(m_{q_{i}}\Sigma_{ij}^{q}R+m_{q_{j}}\Sigma_{ji}^{q*}L\right)\right)q_{j}h   ,\\
L(Hq_{i}q_{j})&=\frac{g}{2M_{W}}\bar{q}_{i}\left(-m_{q_{i}}\frac{c_{\alpha}}{c_{\beta}}\delta_{ij}-\left(\frac{c_{\alpha}}{c_{\beta}}+\frac{s_{\alpha}}{s_{\beta}}\right)\left(m_{q_{i}}\Sigma_{ij}^{q}R+m_{q_{j}}\Sigma_{ji}^{q*}L\right)\right)q_{j}H   ,\\
L(Aq_{i}q_{j})&=-i\frac{I_{q}g}{2M_{W}}\bar{q}_{i}\left(m_{q_{i}}tan\beta\gamma_{5}\delta_{ij}-\left(tan\beta+cot\beta\right)\left(m_{q_{i}}\Sigma_{ij}^{q}R-m_{q_{j}}\Sigma_{ji}^{q*}L\right)\right)q_{j}A     ,\\
L(H^{+}q_{i}q_{j})& =-i\frac{g}{\sqrt{2}M_{W}}\bar{u}_{i}\ \left[m_{d_{j}}tan\beta V_{u_{i}d_{j}}-m_{d_{k}}\left(tan\beta+cot\beta\right)V_{ik}\Sigma_{kj}^{d}\right]R+         ,\\
&\left[-m_{u_{i}}tan\beta V_{u_{i}d_{j}}+m_{u_{k}}(tan\beta+cot\beta)\Sigma_{ki}^{u*}V_{kj}\right]L\} d_{j}H^{+}   ,
\end{split} \label{yuk}
\end{equation}
where $q=u,d$, $I_u,I_d=+\frac{1}{2},-\frac{1}{2}$, $V$ is 4x4 CKM matrix, $\alpha$ is the mixing angle in the CP even neutral Higgs sector and $\tan \beta$ is the ratio between the VEV of $\Phi_h$ and the VEV of $\Phi_l$, i.e. $\tan \beta=\frac{v_h}{v_l}$. Also, $\Sigma^{q}$ are new mixing matrices in the up and the down sectors accordingly, which, for our choice of $\alpha_q,\beta_q$, can be parametrized in the following way~\cite{SBS}:
\begin{equation}
\begin{split}
\Sigma^d & \simeq  \left( \begin{array}{cccc}
0 & 0 &0 &0 \\
0 & 0 &0 &0\\
0 & 0  &\left|\epsilon_b\right|^2 &\epsilon_b^* \\
0 & 0 &\epsilon_b &1-\frac{\left|\epsilon_b\right|^2}{2} \\
 \end{array} \right) \\
\Sigma^u &=  \left( \begin{array}{cccc}
0 & 0 &0 &0 \\
0 & 0 &0 &0\\
0 & 0  &\left|\epsilon_t\right|^2 &\epsilon_t^* \\
0 & 0 &\epsilon_t &1-\frac{\left|\epsilon_t\right|^2}{2} \\
 \end{array} \right) \\
 \end{split}
 \end{equation}
where $\epsilon_b$ and $\epsilon_t$ effectively parametrize the mixing between the third and fourth generation quarks in the down and up sectors, respectively. In the following we will set $\epsilon_b=\frac{m_{b}}{m_{b^\prime }}$ and $\epsilon_t=\frac{m_{t}}{m_{t^\prime  }}$, as natural values representative for these quantities (see \cite{SBS}).
\section{Beyond SM4 decay patterns of $t^\prime  $ and $b^\prime $; a 4G2HDM case}
As mentioned above, in the 4G2HDM the $4^{th}$ generation quarks have a more complex decay pattern that stems from their interactions with the extended Higgs sector. In addition, the SM4 forbidden channels, $t^\prime  \rightarrow Wb^\prime $ and $b^\prime \rightarrow Wt^\prime  $, are no longer in contradiction with the EWPD, and may be kinematically open as well. For simplification, we will assume that $h$ is the only neutral scalar light enough to take part in the decays of $t^\prime  $ and $b^\prime $. We will thus consider the following decay modes:
\begin{enumerate}
\item $t^\prime  \rightarrow ht$ ($b^\prime \rightarrow hb$).
\item $t^\prime  \rightarrow H^+b$ ($b^\prime \rightarrow H^-t$).
\item $t^\prime  \rightarrow Wb$ ($b^\prime \rightarrow Wt$).
\item $t^\prime  \rightarrow Wb^\prime $ ($b^\prime \rightarrow Wt^\prime  $).
\end{enumerate}
We assume the decay mode $t^\prime  \rightarrow H^+b^\prime $ to be kinematically closed taking $M_{H^+}$ to be larger than the mass gap in the fourth generation doublet. 
\\To complete the picture of $t^\prime  $ and $b^\prime $ decays, we need to fully account for the light neutral and charged Higgs decays. For $h$ we have:

\begin{figure}
\begin{center}
	\includegraphics[scale=0.3]
{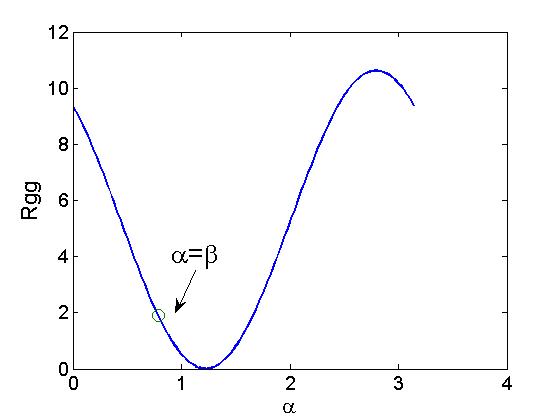}
\end{center}
\caption{$Rgg=\frac{\Gamma \left(h \rightarrow gg \right)_{4G2HDM}}{\Gamma \left(h \rightarrow gg \right)_{SM}} $ as a function of $\alpha$ for 
$m_{t^\prime  }=m_{b^\prime }=450$ GeV and $\tan{\beta}=1$. \label{Rgg} }
\end{figure}

\begin{enumerate}
\item $h\rightarrow \overline{b}b$. This decay mode is prominent for $M_h<130$
GeV.
\item $h\rightarrow W^+W^-,ZZ$. These modes can be suppressed in this model (as in any 2HDM model), due to the $\sin \left(\alpha-\beta\right)$ factor in the $hWW$ and $hZZ$ vertices.
\item $h\rightarrow gg/ \gamma\gamma$. Both channels are loop induced. In the SM4, $h\rightarrow gg$ is the leading decay channel for $M_h<150$ GeV and is about 10 times larger than the SM, while $h\rightarrow \gamma \gamma$ is about 10 times smaller than the SM value \cite{Tait}. On the other hand, in the 4G2HDM, $h\rightarrow gg$ could be suppressed in some areas of the parameter space, as a result of a destructive interference between the heavy fermion loops due to extra factors in the corresponding Yukawa terms which depend on the angles $\alpha$ and $\beta$ (see Eq. \ref{yuk}). This would lead to either a suppression or enhancement of light Higgs production by gluon fusion as a function of $\alpha$, as is demonstrated in Figure~\ref{Rgg}. It is obvious that, even for the "popular" choice of $\alpha=\beta$, the enhancement of Higgs production is only by a factor of ~2, which is almost 5 times smaller than the corresponding enhancement in the SM4. This is an example of how an extended Higgs sector may suppress the Higgs production signal (see also \cite{Val4,Gunion}), providing a possible explanation of the absence of an observed signal \cite{CMS_H_limit}. Similarly, $h\rightarrow \gamma \gamma$ could be enhanced or suppressed in the 4G2HDM with respect to the SM4 or the SM \cite{Gunion,Bern}. These issues are beyond the scope of this work and therefore will not be discussed here.
\item $h\to \nu_{\tau'}\nu_{\tau'}/\tau'\tau'/\overline{t}t$. For the purposes of this work, we assume that $M_h<2m_{\nu_{\tau'}}/2m_{\tau'}/2m_t$ allowing us to ignore these channels.
\end{enumerate}
The $H^+$ decays are the following:
\begin{enumerate}
\item $H^+ \rightarrow tb$
\item $H^+ \rightarrow t^\prime   b$ 
\item $H^+ \rightarrow hW^+$
\item $H^+ \rightarrow \tau \nu_{\tau}$ and $H^+ \rightarrow \tau' \nu_{\tau'}$.
\end{enumerate} 
The decay modes for $H^+$ that involve the first two generations are much smaller than the above ones, due their light masses. 
Combining all these channels to a cascade decay, we are left with several topologies for the $t^\prime  $ signatures, with different decay chains for each possibility:
\begin{enumerate}
\small
\item $t^\prime   \rightarrow W+b$ (SM4-like)
\item $t^\prime   \rightarrow W+3b$:
\begin{itemize}
\small
\item $t^\prime  \rightarrow ht \rightarrow \left[\overline{b}b \right]_h \left[W^+b \right]_{t}$
\item $t^\prime  \to H^+b \rightarrow \left[t \overline{b} \right]_{H^+} b \rightarrow  \left[\left(W^+b \right)_t \overline{b} \right]_{H^+} b$
\item $t^\prime   \rightarrow W^+b^\prime  \rightarrow   W^+ \left[hb \right]_{b^\prime } \rightarrow W^+  \left[\left(\overline{b}b\right)_hb \right]_{b^\prime } $
\end{itemize}
\item $t^\prime   \rightarrow 3W+b$
\begin{itemize}
\item  $t^\prime  \rightarrow ht\rightarrow \left[W^+W^- \right]_h \left[W^+b \right]_t$
\item $t^\prime   \rightarrow H^+b \to \left[hW^+\right]_{H^+} b \rightarrow \left[\left(W^+W^-\right)_hW^+\right]_{H^+} b$
\item $t^\prime   \rightarrow W^+b^\prime  \rightarrow   W^+ \left[hb \right]_{b^\prime } \rightarrow W^+  \left[\left(W^+W^-\right)_hb \right]_{b^\prime } $
\end{itemize} 
\end{enumerate}

While for $b^\prime $ decays we have:
\begin{enumerate}
\item $b^\prime  \rightarrow 3b$
\begin{itemize}
\item $b^\prime  \rightarrow hb \rightarrow \left[\overline{b}b \right]_hb$
\end{itemize}
\item $b^\prime  \rightarrow 2W+b$ (SM4-like)
\begin{itemize}
\item $b^\prime  \rightarrow hb\rightarrow \left[W^+W^-\right]_hb$
\item $b^\prime  \rightarrow W^-t \rightarrow \left[W^+b \right]_t W^- $ (SM4- Like)
\item $b^\prime  \rightarrow W^-t^\prime   \rightarrow \left[W^+b \right]_{t^\prime  } W^- $
\end{itemize}
\item $b^\prime  \rightarrow 4W+b$
\begin{itemize}
\item $b^\prime  \rightarrow H^-t \rightarrow \left[hW^-\right]_{H^-} \left[W^+b\right]_t \rightarrow \left[\left(W^+W^-\right)_hW^-\right]_{H^-} \left[W^+b\right]_t $
\item $b^\prime  \rightarrow  t^\prime  W^- \rightarrow \left[ht \right]_{t^\prime  } W^- \rightarrow  \left[\left(W^+W^- \right)_h \left(W^+b \right)_t \right]_{t^\prime  } W^-$
\item $b^\prime  \rightarrow  W^- \rightarrow \left[H^+ b \right]_{t^\prime } W^- \rightarrow  \left[\left(hW^+ \right)_{H^+} b \right]_{t^\prime } W^- \rightarrow \left[\left(\left\{W^+W^-\right\}_hW^+ \right)_{H^+} b \right]_{t^\prime } W^-$
\end{itemize}
\item $b^\prime  \rightarrow 2W+3b$

\begin{itemize}
\item $b^\prime  \rightarrow H^-t \rightarrow \left[hW^-\right]_{H^-} \left[W^+b\right]_t \rightarrow \left[\left(\overline{b}b\right)_hW^-\right]_{H^-} \left[W^+b\right]_t $
\item $b^\prime  \rightarrow  t^\prime  W^- \rightarrow \left[ht \right]_{t^\prime } W^- \rightarrow  \left[\left(\overline{b}b \right)_h \left(W^+b \right)_t \right]_{t^\prime } W^-$
\item $b^\prime  \rightarrow  t^\prime  W^- \rightarrow \left[H^+ b \right]_{t^\prime } W^- \rightarrow  \left[\left(hW^+ \right)_{H^+} b \right]_{t^\prime } W^- \rightarrow \left[\left(\left\{\overline{b}b\right\}_hW^+ \right)_{H^+} b \right]_{t^\prime } W^-$
\item  $b^\prime  \rightarrow  t^\prime  W^- \rightarrow \left[H^+ b \right]_{t^\prime } W^- \rightarrow  \left[\left(t\overline{b} \right)_{H^+} b \right]_{t^\prime } W^- \rightarrow \left[\left(\left\{W^+b\right\}_t\overline{b} \right)_{H^+} b \right]_{t^\prime } W^-$
\end{itemize} 
\end{enumerate}

The decay chains not listed here, e.g. decay chains with Z bosons or fourth generation leptons, may further complicate the phenomenology of $t^\prime $ and $b^\prime $ and will not be considered in this work. 

The above $t^\prime $ and $b^\prime $ decay channels give rise to a set of new signatures for $t^\prime $ and $b^\prime $ pair production at the LHC which we will generically denote by  
	$pp\rightarrow t^\prime \overline{t^\prime }/b^\prime \overline{b^\prime }\rightarrow n_WW+n_bb$, 
with $n_W$ and $n_b$ being the number of $W$ and $b$ jets in the event, respectively.
In particular, the $t^\prime  t^\prime $ signatures are 2$W$+2$b$ (SM4- like), 2$W$+6$b$, 6$W$+2$b$, 2$W$+4$b$, 4$W$+2$b$ and 4$W$+4$b$. As was mentioned earlier, for these signatures there is an excess of jets, leptons and $b$-jets with respect to the SM and the SM4 signals, as is demonstrated in Figure~\ref{nj_bb}. 

We stress again that, in general, these signatures are not exclusive to the 4G2HDM; they can appear in any model consisting of new (neutral or charged) scalars in the reasonable scenario that they interact mostly with the $3^{rd}$ and $4^{th}$ generation fermions.

\section{Results: semileptonic channel}

 For the semileptonic channel ($pp\rightarrow t^\prime \overline{t^\prime }\rightarrow 1\ell+nj+\rlap{ /}{E}_T$), we suggest a general reconstruction method that is based on \cite{SGA} and test the extent to which the new signatures are already excluded by the "old" CMS analysis (which gives $m_{t^\prime } > 450$ GeV for the SM4) \cite{CMS_limit}. We focus our study bellow on the new $pp \to t^\prime \overline{t^\prime }\to 2W+6b$ or $6W+2b$ signals, and show that, in those cases, $t^\prime $ can be detected at the LHC using the excess of jets and $b$-jets mentioned above. The analysis is performed for two choices of parameter space in the 4G2HDM for which $Br \left(t^\prime  \to th \right)\sim{\cal O}\left(1\right) $, but arguably, due to the fact that no care is given to the specifics of each decay chain, our conclusions hold for any other model or parameter space that results in the same signature (in the previous section, we showed that for each signature there are several decay chains in the 4G2HDM with different particles in the intermediate states but with the same final state).
 
We use Madgraph/MadEvent \cite{MADGRAPH} to generate the signal and two $\int Ldt=1$ fb$^{-1}$  sets of background events: $W$+jets and  $t\overline{t}$+jets, with $K$-factors of 1.5 for the signal and the $t\overline{t}$+jets background \cite{Hold_tp}, and 1.3 for $W$+jets background \cite{D0_limit}. We use MLM parton-jet matching method \cite{MLM} for the background with $p_{Tmin}=100$ GeV for the $t\overline{t}$+jets and $p_{Tmin}=150$ GeV for $W$+jets (see \cite{Hold_tp}). We use BRIDGE \cite{BRIDGE} for the decay of the new particles and Pythia \cite{Pythia} for the decay of the SM particles, shower, fragmentation and hadronizations. We use PGS \cite{PGS} with the LHC card for the detector simulation.

We look for $1\ell+nj+\rlap{ /}{E}_T$ signatures, requiring exactly one muon or electron with $p_T>20$ GeV, $\rlap{ /}E_T>20$ GeV, and at least four jets with $p_T>120,90,35,35$ GeV and $\left|\eta\right|<2.5$

For the signal, we choose $m_{t^\prime }=350,400,450$ GeV, $\tan\beta=1$ , $V_{34}=0.1$, and two different values for the Higgs mass, $M_h=130,170$ GeV$^{\left[1\right]}$  \footnotetext[1]{Although $M_h \approx 125 GeV$ seems more realistic in view of the recent hints from the LHC, we fix $M_h=130, 170$ GeV for demonstration purposes.} , while the other scalars are assumed to be heavier and thus play no role in the decays. For both values of $M_h$ the main decay mode is $t^\prime  \rightarrow ht$ with a branching ratio of $\sim0.9$. For $M_h=170$ GeV, $h$ decays mainly to $W^+W^-$, while $h\rightarrow b\overline{b}$ for $M_h=130$ GeV; this leaves us with the two aforementioned signatures: $6W+2b$ and $2W+6b$, respectively. 
\begin{figure}[t]
\begin{center}
	\includegraphics[scale=0.6]
{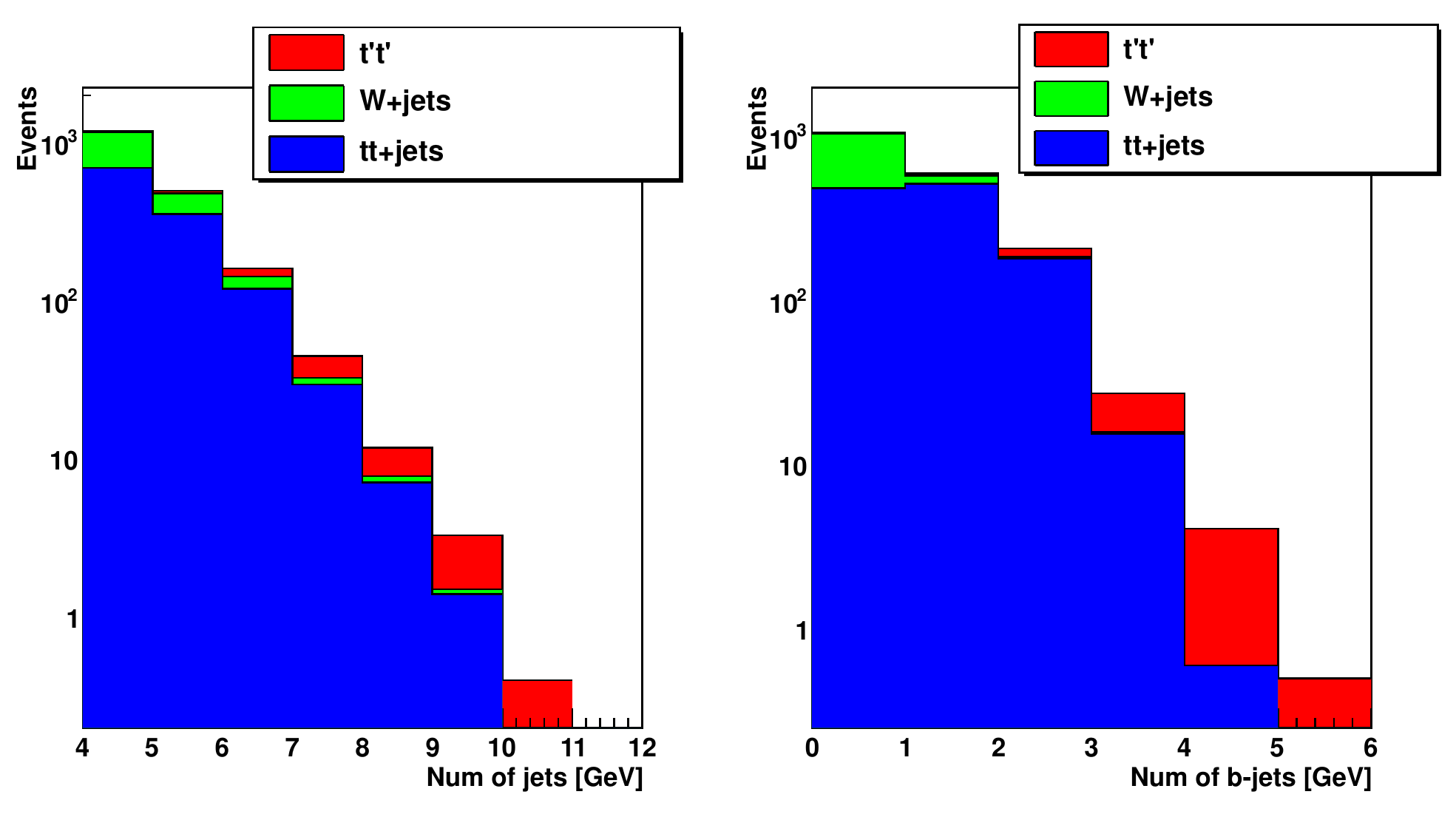}
\end{center}
\caption{Number of jets (left) and number of $b$-jets (right) histograms for the $pp\rightarrow t^\prime \overline{t^\prime }\rightarrow 2W+6b$ signature(red) with 
$m_{t^\prime }=450$ GeV. Also shown are the backgrounds from $W$+jets (green) and $t\overline{t}$+jets (blue) for a set of 7 TeV LHC events with $\int Ldt=1$ fb$^{-1}$, in the semileptonic channel ($1\ell+nj+\rlap{ /}{E}_T$).  \label{nj_bb}}
\end{figure}
\begin{figure}[t]
\begin{center}
	\includegraphics[scale=0.5]
{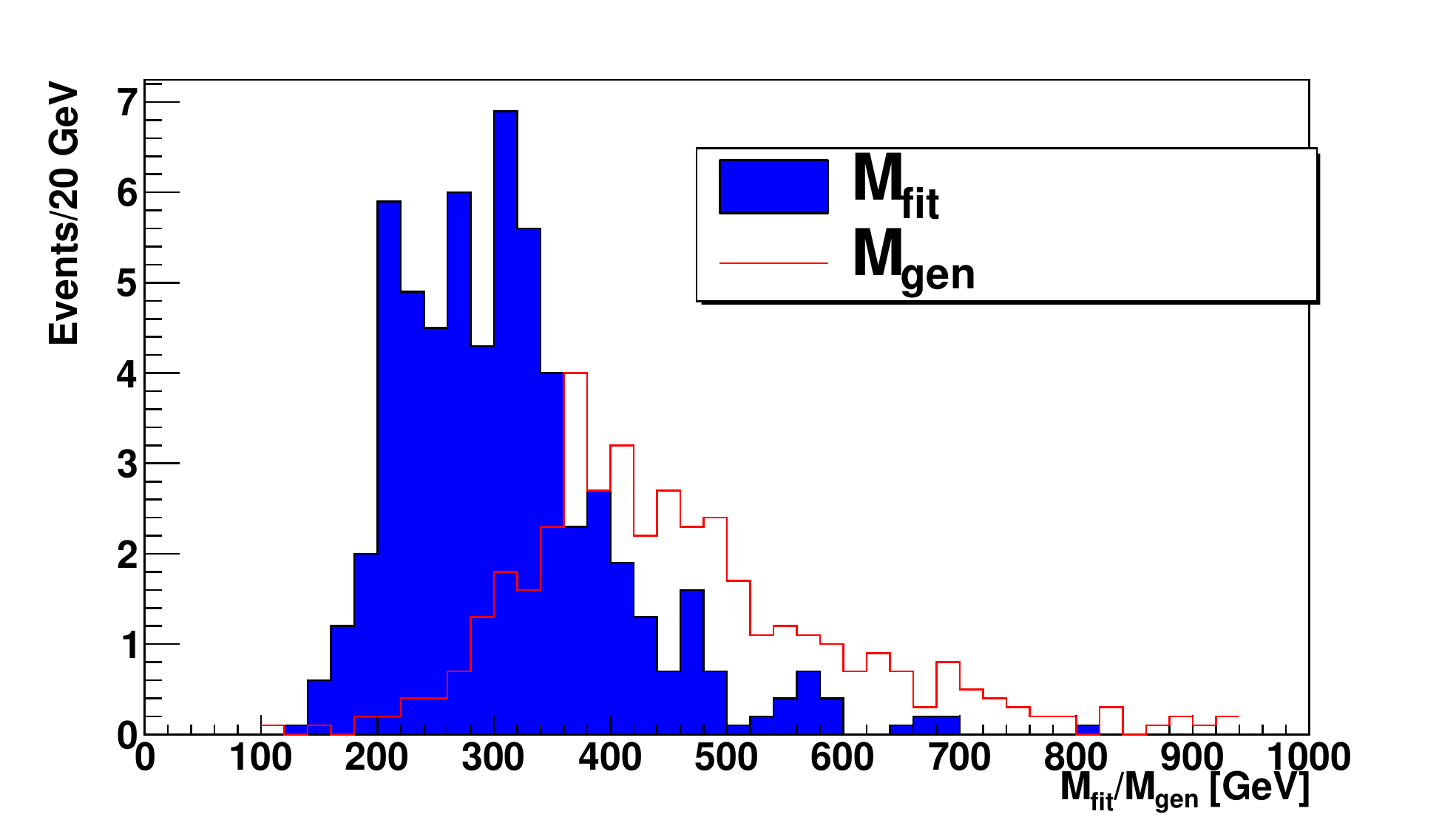}
\end{center}
\caption{ Comparison between $M_{fit}=m\left(l\nu b\right)=m\left(q\overline{q}b\right)$ (the reconstructed $t^\prime $ mass using the CMS method, see Eq. \ref{CMS_Method}) in blue and $M_{gen}=m\left(Left\;  Side \right)=m\left(Right\; Side \right)$ (the reconstructed $t^\prime $ mass using our method, see  Eq.\ref{Our_Method}  ) in red, for the $pp\rightarrow t^\prime \overline{t^\prime }\rightarrow 2W+6b$ signature with $m_{t^\prime }=450$ GeV at the LHC (7 Tev) with $\int Ldt=1$ fb$^{-1}$, in the semileptonic channel ($1\ell+nj+\rlap{ /}{E}_T$). See also text.\label{gen_fit}}
\end{figure}
\begin{figure}[t]
\begin{center}
	\includegraphics[scale=0.5]
{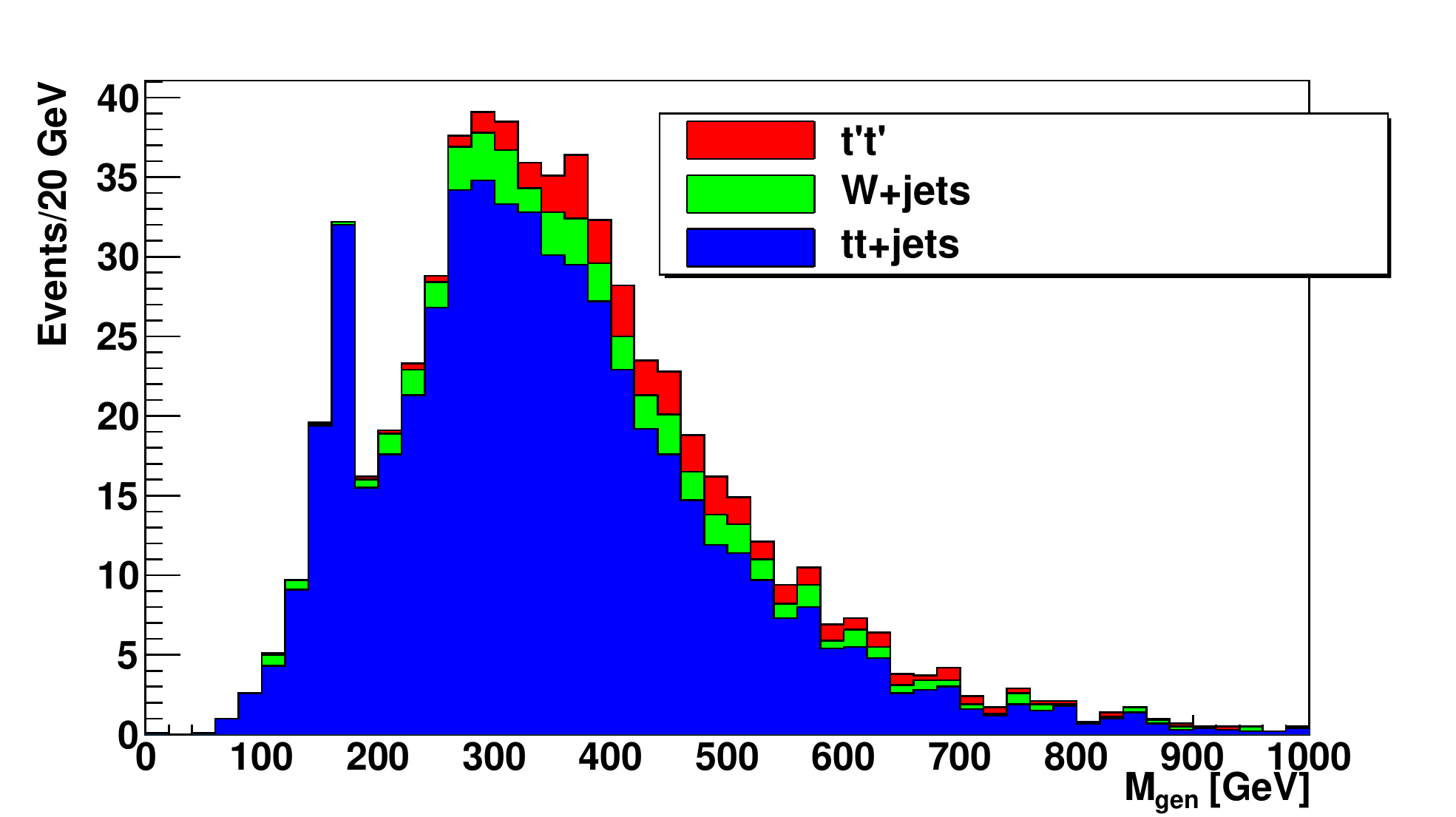}
\end{center}
\caption{$M_{gen}$ for $pp\rightarrow t^\prime \overline{t^\prime }\rightarrow 2W+6b$ with $m_{t^\prime }=450$ GeV (red) together with $W$+jets (green) and $t\overline{t}$+jets backgrounds (blue) at the LHC (7 Tev) with $\int Ldt=1$ fb$^{-1}$, in the semileptonic channel ($1\ell+nj+\rlap{ /}{E}_T$).\label{Mgen_bkg}}
\end{figure}
\begin{figure}[t]
\begin{center}
	\includegraphics[scale=0.5]
{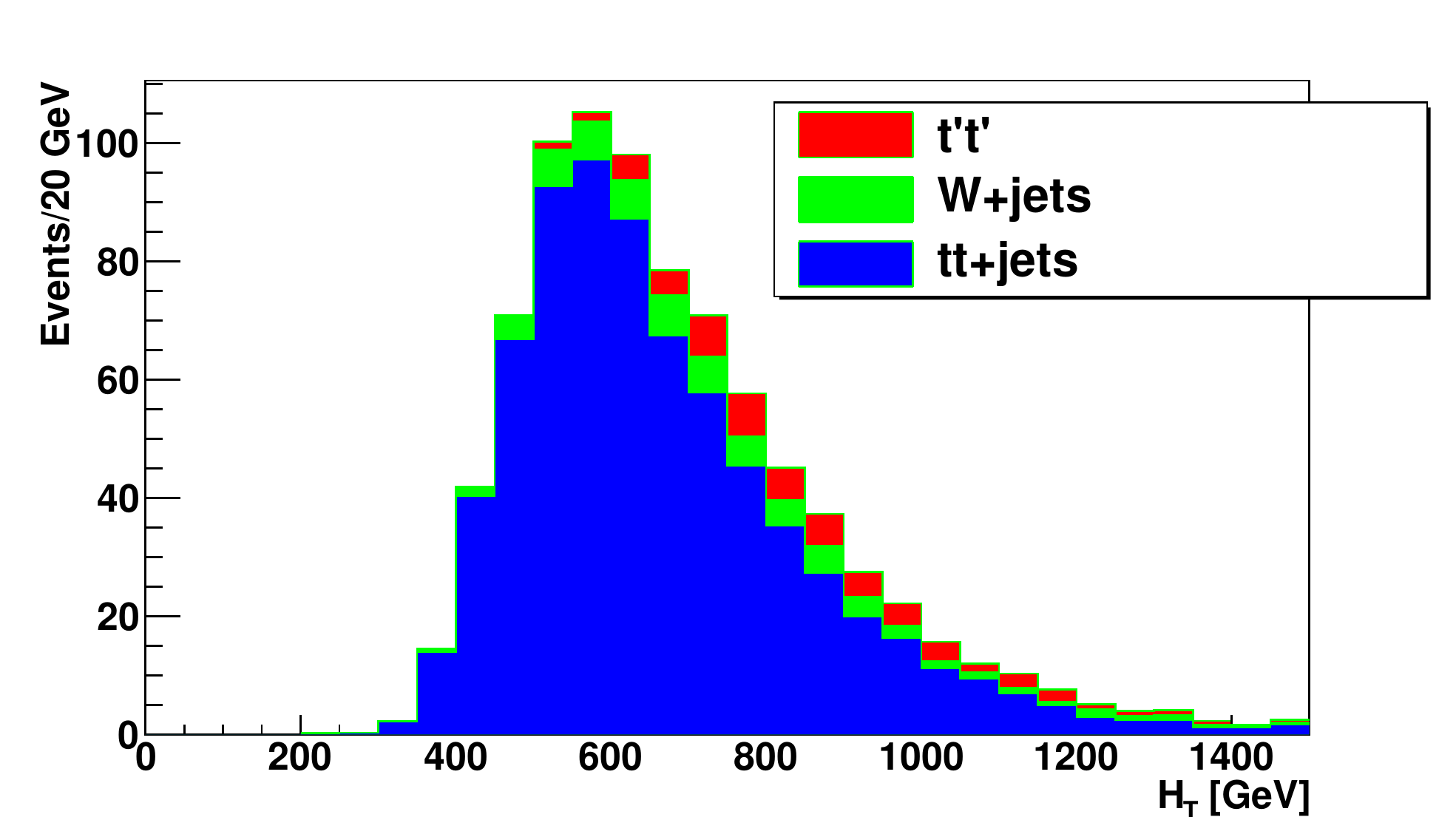}
\end{center}
\caption{$H_{T}$ for $pp\rightarrow t^\prime \overline{t^\prime }\rightarrow 2W+6b$ with $m_{t^\prime }=450$ GeV (red) together with $W$+jets (green) and $t\overline{t}$+jets backgrounds (blue) at the LHC (7 Tev) with $\int Ldt=1$ fb$^{-1}$, in the semileptonic channel ($1\ell+nj+\rlap{ /}{E}_T$). \label{HT}}
\end{figure}
 First we demonstrate in Figure~\ref{nj_bb} the expected excess of jets and $b$-jets for the 6$b$+2$W$ signature. We would not use the jet and $b$-jet distribution as a signal, as that might underestimate the background, especially the high multiplicity one ($t\overline{t} +nj$ with $n>2$). Instead, we use the demonstrated excess as a basis for one of the cuts in the selection of the signal. When enough low multiplicity background events survive the cut, the addition of the high multiplicity background events will not play a dramatic role in the $S/B$ ratio and will not change our conclusions.
 
 For the same signature, we compare $M_{fit}$ (the reconstructed mass with the SM4 hypothesis) and $M_{gen}$ (the general reconstruction mentioned earlier, see Eqn.~\ref{Our_Method}) in Figure~\ref{gen_fit}, while in Figure~\ref{Mgen_bkg} we plot the distribution of $M_{gen}$ for the signal (2$W$+6$b$) and background events. We can see in Figure~\ref{gen_fit} that the peak of the distribution of $M_{gen}$ is roughly at $m_{t^\prime }$, while for $M_{fit}$ the peak is misplaced, being substantially lower. We also note that the distribution of $M_{gen}$ in the new signatures is much broader around the peak than the distribution of $M_{fit}$; this is due to the massive combinatorial background coming from the high number of jets in the final state. In Figure~\ref{HT} we plot the distribution of $H_T$ for the signal and the background events for the same signature.

\begin{table}
\begin{tabular} {|p{5cm} p{2cm} c c c ||c|c|}
\hline \hline
 Process & without cuts &$n_{j+bj}$ & $M_{gen}$ & $H_T$ & all cuts combined & $\frac{S}{\sqrt{B}}$\\ \hline
 $t\overline{t}+$jets     &1206 &179 &536 &678& $93$ &  \\ 
 $W+$jets     &626 &9 &353 & 443& $5$   &\\  \hline 
  2$W$+6$b$  $m_{t^\prime }=350$ GeV   &172 &99 &93 &138& 56&  5.66\\ 
  2$W$+6$b$  $m_{t^\prime }=400$ GeV   &119 &74 &69 &104& 45 &  4.55\\ 
    2$W$+6$b$  $m_{t^\prime }=450$ GeVe   &70 &46 &43 &65& 29& 2.93\\  \hline
 6$W$+2$b$  $m_{t^\prime }=350$ GeV   &168 &80 &87 &135& 46&4.65 \\ 
 6$W$+2$b$  $m_{t^\prime }=400$ GeV\   &113 &58 &60 &99& 34& 3.43\\ 
 6$W$+2$b$  $m_{t^\prime }=450$ GeV &77 &43 &40 &74& 23&2.32 \\
 \hline
\end{tabular}
\caption{Number of events after the selection (in the second column), after each of the three cuts: $n_{j+bj}\equiv n_{jets}+n_{b-jets}>6$, $M_{gen}>300$ GeV
, $H_T>600$ GeV ($3^{rd}- 5^{th}$ columns) and all of them combined ($6^{th}$ column). $\frac{S}{\sqrt{B}}$ for each of the new signatures at the LHC (7 Tev) with $\int Ldt=1$ fb$^{-1}$ is given in the seventh column. \label{table_sens} }
\end{table}

Based on the above, we test the sensitivity of detection in our method for the two signatures (6$W$+2$b$ and 2$b$+6$W$) for the 3 $t^\prime $ mass values, $m_{t^\prime }=350,400,450$ GeV, using the following cuts:
\begin{enumerate}
\item $M_{gen}>300$ GeV
\item $H_T>600$ GeV
\item $n_{j+bj}\equiv n_{jets}+n_{b-jets}>6$
\end{enumerate}

The results are summarized in Table~\ref{table_sens}. Wee see that the leading background is $t\overline{t}$+jets, as $W+$jets almost vanishes when imposing $n_{jets}+n_{b-jets}>6$. For $m_{t^\prime }=350$ GeV we get a statistical significance of $\frac{S}{\sqrt{B}}\sim 4.7-5.7$, depending on the signature. For higher values of $m_{t^\prime }$ the statistical significance drops, e.g., to the level of $\sim 2.3-2.9$ for the two signatures when $m_{t^\prime }=450$ GeV . In contrast, applying the "standard" CMS method (i.e. using $M_{fit}$ instead of $M_{gen}$ and requiring that at least one of the jets is tagged as a $b$-jet instead of the $n_{j+bj}$ cut), we get $\frac{S}{\sqrt{B}}\sim 2.6$ for $m_{t^\prime }=350$ GeV and $\frac{S}{\sqrt{B}}\sim 1.7$ for $m_{t^\prime }=450$ GeV.

 We proceed to test the consistency of the CMS exclusions results in the semileptonic channel (which exclude $t^\prime  \to Wb$ as is the case in the SM4) with the existence of a BSM4 signal in the current data. We recall that the CMS analysis is based on a 2d ($M_{fit}$ and $H_T$) likelihood fit to the data under the hypothesis of signal (SM4) + background and the hypothesis of background only, and a CLs  method \cite{CLs1,CLs2} for the computation of the exclusion limits. They found that 
$m_{t^\prime }>450$ GeV at 95\% CL (for the SM4 case) \cite{CMS_limit}. 
\begin{figure}[t]
\begin{center}
	\includegraphics[scale=0.6]
{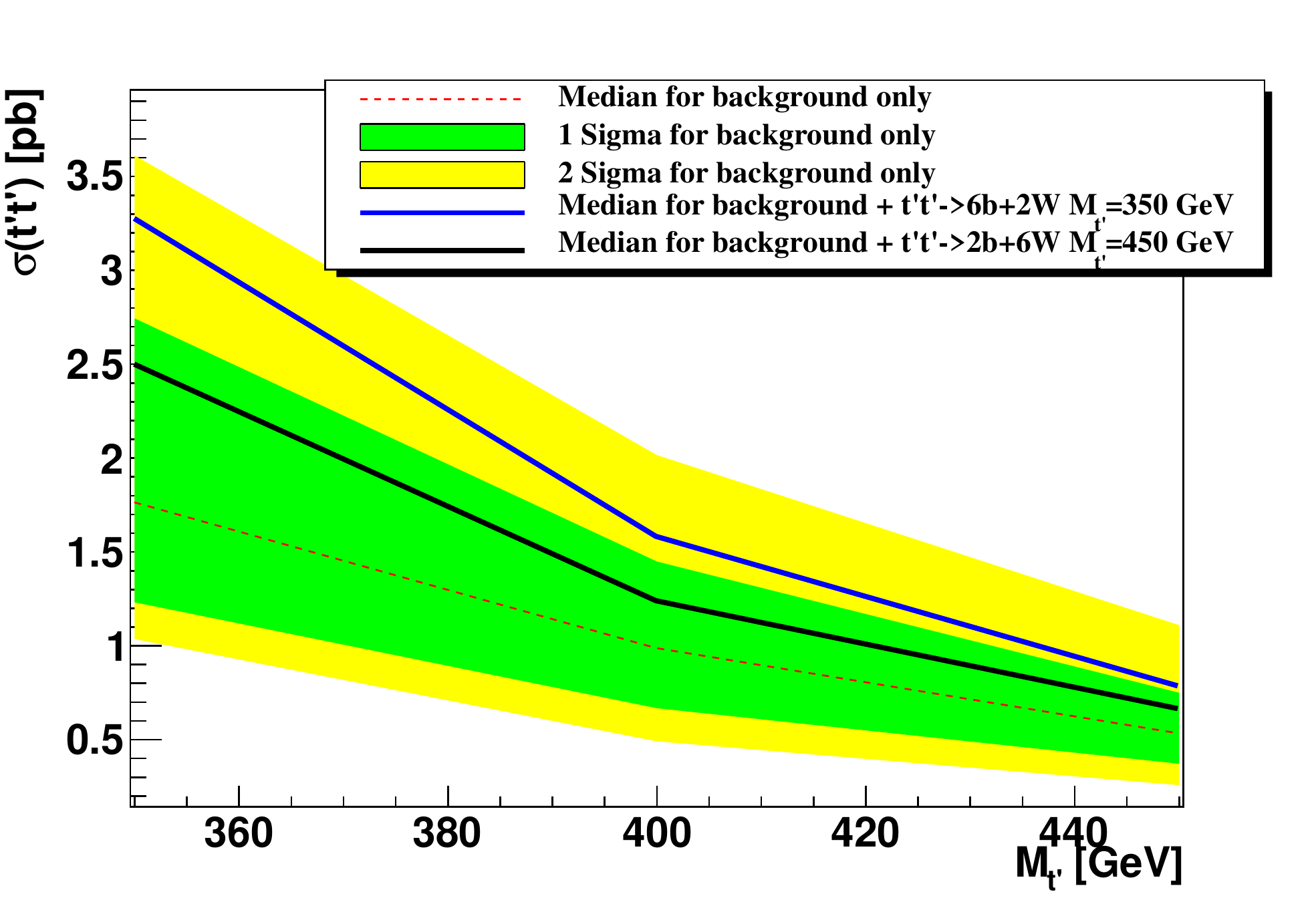}
\end{center}
\caption{The 95\% CL exclusion plot distribution on the $t^\prime $ mass assuming the SM4 signature in the semileptonic channel ($1\ell+nj+\rlap{ /}{E}_T$) . For the case of background only, the red doted line is the median and the yellow and green bands are the $\pm1$ and $\pm2$ standard deviations accordingly. The black line is the median for background + $t^\prime  t^\prime  \to 2b+6W$ with $m_{t^\prime }=450$ GeV and the blue line is the median for background + $t^\prime  t^\prime  \to 6b+2W$ with $m_{t^\prime }=350$ GeV. The results for the rest of the signatures with $m_{t^\prime }=350-450$ GeV lie between these two lines. \label{Limit}}
\end{figure}

 We perform a simplified computation of what would the CMS exclusion limit be for $m_{t^\prime }$ in the case of background only and in the case of background+BSM4. To do so, we generate 1000 LHC events ($\int Ldt=1$ fb$^{-1}$) for each case. We then use the ratio $L=L_b/L_{s+b}$ of the likelihoods to the signal (SM4) + background hypothesis ($L_{s+b}$) and to the background hypothesis ($L_b$), as the test statistic. We naively assign a 20\% systematic uncertainty on the background cross-section, minimizing the likelihood with respect to this parameter in each case. We calculate the likelihood ration ($L$) for each simulated signature, and determine the probability to observe a higher value of $L$ for the background only hypothesis ($CL_b$) and the signal+background hypothesis($CL_{s+b}(\sigma$)). The 95\% CL limit on the signal cross-section is then the value of the signal cross-section $\sigma$, for which
\begin{equation}
CL_s=\frac{CL_{s+b}(\sigma)}{CL_b}=0.05.
\end{equation}
Finally, we calculate the distribution of the 95\% CL limit on the signal cross-section for background only and for background+BSM4 hypothesis, and plot in Figure~\ref{Limit} the exclusion limits for background only, and for background with two chosen BSM4 signatures for $t^\prime $ pair production: $t^\prime t^\prime  \to 2b+6W$ with $m_{t^\prime }=450~$ GeV  and $t^\prime t^\prime  \to 6b+2W$ with $m_{t^\prime }=350$ GeV. The exclusion limits for all the other BSM4 signatures that we've checked (6$b$+2$W$,2$W$+6$b$ with $m_{t^\prime }=350,400,450$ GeV) are between these two limits for the chosen signatures. We observe that for $m_{t^\prime }=350$ GeV, the median of the exclusion limit distribution for the BSM4 signatures is within 2 standard deviations of the expected limit for the background only case, and for $m_{t^\prime }=450$ GeV it is fully within 1 standard deviation of the expected limit for the background only case. Thus, the discrepancy between the CMS observed exclusion limit, which is close to the expected one, see \cite{CMS_limit} (the expected limit in our naive calculation is denoted by the red dotted line in Figure~\ref{Limit}), and the aforementioned BSM4 scenarios is less than $2\sigma$ for $m_{t^\prime }\gsim 350$ GeV, and 1-2$\sigma$ for $m_{t^\prime }=350$ GeV. We, therefore, go on to conclude that BSM4 scenarios with $m_{t^\prime }>350$ GeV, can be consistent with the current semileptonic data and therefore the limit on the $t^\prime $ mass for the BSM4 case should be around $350$ GeV in the semileptonic channel.

\section{Results: the dilepton channel}
The most stringent bounds to this date on the $t^\prime $ mass (557 GeV) are set by the CMS, analyzing the dilepton channel ($pp\rightarrow t^\prime \overline{t^\prime } \rightarrow \ell^+\ell^- \nu \overline{\nu} b \overline{b}$) for $t^\prime$  pair production and decay at the LHC \cite{CMS_limit_dilepton}. The signal in this channel is relatively rare (about 11\% of the $t^\prime \overline{t^\prime }$ events) but clean due to the presence of the two leptons that suppresses the SM background. The suppression of the background is further promoted by requiring that exactly two of the jets in each selected event are tagged as $b$-jets. The kinematic variable that is used to discriminate the signal from the huge $t\overline{t}$ background is the invariant mass of one of the leptons and one of the $b$-jets, $M^{min}_{lb}$, where the configuration with the lowest invariant mass is chosen. In the case of $t\overline{t}$ production this invariant mass is bounded from above by the top mass (170 GeV), while for the $t^\prime \overline{t^\prime }$ signal, there should be a large portion of events giving $M^{min}_{lb}>170$. This leaves the region above this mass (named the "Signal Region") almost background-free (except mostly misidentification background) and completely signal-dominated. The number of events in the Signal Region is therefore used for the limit calculation.

When the new textures of $t^\prime $ decay with higher multiplicities of leptons and $b$-jets are considered, the dominance of the signal for $M^{min}_{lb}>170$ GeV may no longer be realized, due to the combinatorial background and the altered kinematic shape of the events. We try to address this issue using a simulation similar to the one in the previous section, with different selection rules. We generate $t^\prime \overline{t^\prime }$ events with $5$ fb$^{-1}$ for the SM4-like 2$W$+2$b$ signature and for the new 2$W$+6$b$ signature, choosing $m_{t^\prime }=350$ GeV and selecting events with at least two leptons and exactly two $b$-jets, with the kinematic cuts that were used in the CMS analysis \cite{CMS_limit_dilepton}. Due to the fact that the $b$-tagging efficiency in PGS \cite{PGS} is not well calibrated with the efficiency of the TCHEM method of $b$-tagging \cite{TCHEM} that was used in the CMS analysis, we get a lower number of events for the SM4-like signature than the number of events reported in \cite{CMS_limit_dilepton}. We thus use a higher efficiency to roughly simulate the correct number of events in the Signal Region. Thus, our results below should be considered only as an illustration of the effect of BSM4.
\begin{figure}[t]
\begin{center}
	\includegraphics[scale=0.6]
{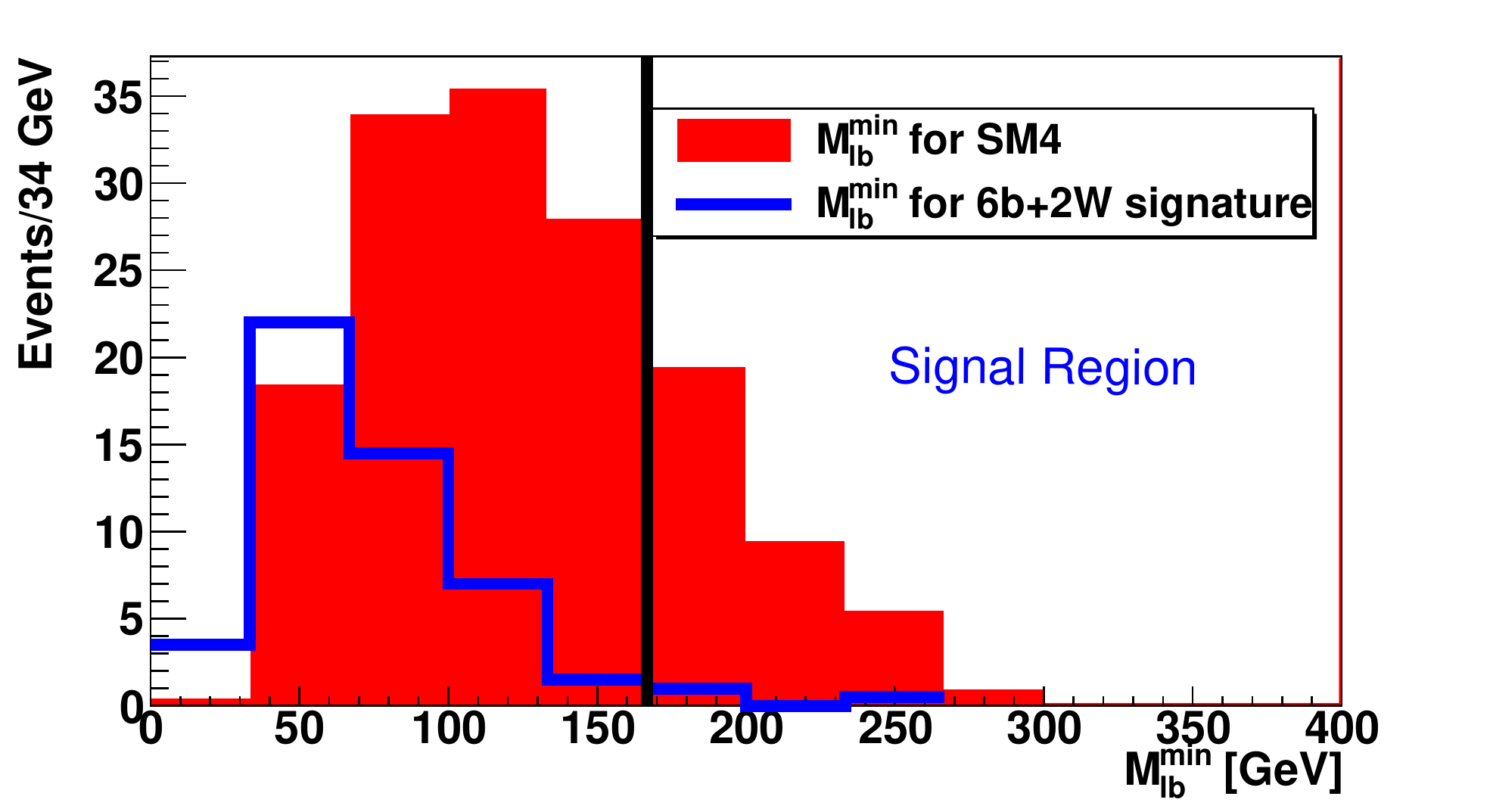}
\end{center}
\caption{$M^{min}_{lb}$ for the SM4-like $pp\rightarrow t^\prime \overline{t^\prime }\rightarrow 2W+2b$ signature (red) and for the BSM4 $pp\rightarrow t^\prime \overline{t^\prime }\rightarrow 2W+6b$ signature(blue) with 
$m_{t^\prime }=350$ GeV for a set of 7 TeV LHC events with $\int Ldt=5$ fb$^{-1}$ in the dilepton channel. The black line is approximately at the top mass and the region to the right of the line is the signal region. \label{Mlb_dilepton}}
\end{figure}

The results are presented in Figure~\ref{Mlb_dilepton}. We first note, that the total number of selected events is lower for the 2$W$+6$b$ than for the SM4 case. This is due to the fact that there is a higher number of $b$ jets (6) in 2$W$+6$b$ events, thus, for high enough $b$-tagging efficiency (above 0.5) there is a lower chance of tagging exactly 2 of them. In the signal region, the number of the events for the 2$W$+6$b$ signature is negligible- about 1/10 of the events in SM4 case. In fact, for $m_{t^\prime }>350$ GeV, the number of 2$W$+6$b$ events in the Signal Region is even lower due to the lower production cross-section. We therefore conclude that the current stringent limit on the $t^\prime $ mass (557 GeV) in the dilepton channel is completely irrelevant for some BSM4 scenarios, for $m_{t^\prime }>350$ GeV (we recall that $m_{t^\prime }=350$ GeV is the lowest mass considered in \cite{CMS_limit_dilepton}).

\section{Summary}
We have explored new detection strategies for the $t^\prime $ that hold in the general case, where new decay patterns (beyond the SM4) for $t^\prime $ emerge, e.g. $t^\prime  \to th$. In such cases, new  $t^\prime $ signals, such as $pp\to \overline{t^\prime } t^\prime  \to 6W+2b $ and $pp\to \overline{t^\prime } t^\prime  \to 2W+6b $ that we have studied here may dominate. 
As a representative framework for BSM4 we chose the 4G2HDM of \cite{SBS}, which is a variation of a 2HDM framework with 4 generations. We then studied its implications for $t^\prime $ phenomenology at the LHC, keeping the analysis as general as possible and, therefore, model-independent to a large extent.

We have shown that if the new $t^\prime $ decay channels lead 
to $pp\to \overline{t^\prime } t^\prime  \to 2W+6b$ or to $pp\to \overline{t^\prime } t^\prime  \to 6W+2b$, then
the current limits on the $t^\prime $ mass (450/557 GeV for the semileptonic and the dilepton channel in the SM4 case) do not apply and a different detection strategy is required. 
For the semileptonic channel, we estimate the current limit on the $t^\prime $ mass in the BSM4 case to be around 350 GeV, demonstrating that a general reconstruction method (without the assumption of a specific topology of each event) is successful in pushing the limit upwards. 

In the dilepton channel, we find the CMS limit to be completely irrelevant for the mass range in the analysis (350-600 GeV) for several BSM4 scenarios. This includes the case that the mass of the lightest Higgs is around 130 GeV, which may be consistent with what ATLAS and CMS see in the Higgs searches. We leave for a future study the constraints on new scenarios resulting from the $b\prime$ searches at the LHC  \cite{Atlas_bprime,CMS_bprime}.

\bigskip
{\bf Acknowledgments:}
The authors would like to thank Amarjit Soni for useful discussions.  
GE thanks P.Q. Hung for helpful discussions and SBS acknowledges research support from the Technion. MG thanks Iftah Galon for his help with the numerical analysis.

\bigskip

After completing our work we received a preprint by K.~Rao and D.~Whiteson, (arXiv:1204.4504 [hep-ph]) in which ideas similar to ours
are expressed.

\end{document}